\documentclass[11pt,titlepage]{article}
\renewcommand{\to}{\rightarrow}
\newcommand{\beq}{\begin{equation}}
\newcommand{\eeq}{\end{equation}}
\newcommand{\bea}{\begin{eqnarray}}
\newcommand{\eea}{\end{eqnarray}}
\def\ln{{\rm ln}}
\begin{document}
\thispagestyle{empty}
\begin{titlepage}
\addtolength{\baselineskip}{.7mm}
\thispagestyle{empty}
\begin{flushright}
DFTT 70/99\\
\end{flushright}
\vspace{10mm}
\begin{center}
{\Large{\bf Three-dimensional QCD in the adjoint representation
and random matrix theory
}}\\[15mm]
{
\bf 
U.~Magnea 
} \\
\vspace{5mm}
{INFN \\
Via P. Giuria 1, I-10125 Torino, Italy \\
blom@to.infn.it
}\\[6mm]
\vspace{13mm}
{\bf Abstract}\\[5mm]
\end{center}

In this paper we complete the derivations of finite volume partition 
functions for QCD using random matrix theories by calculating the 
effective low-energy 
partition function for three-dimensional QCD in the adjoint
representation from a random matrix theory with the same global symmetries. 
As expected, this case corresponds to 
Dyson index $\beta =4$, that is, the Dirac operator can be written in 
terms of real quaternions. After discussing the issue of defining 
Majorana fermions in Euclidean space, 
the actual matrix model calculation turns out to be simple.
We find that the symmetry breaking pattern is $O(2N_f) \to O(N_f) \times
O(N_f)$, as expected from the correspondence between symmetric
(super)spaces and random matrix universality classes found by Zirnbauer.
We also derive the first Leutwyler--Smilga sum rule.

\end{titlepage}
\newpage
\renewcommand{\theequation}{\arabic{section}.\arabic{equation}}
\renewcommand{\thefootnote}{\fnsymbol{footnote}}
\setcounter{footnote}{0}

\section{Introduction}
\label{sec-Intro}

The recent strong interest in the low-energy limit of the QCD spectrum
originates in the Banks-Casher relation \cite{BC}

\begin{equation}
\Sigma = \frac{\pi \rho(0)}{V}
\end{equation}

relating the density of eigenvalues of the Dirac operator at the origin, 
$\rho(0)$, to the order parameter for spontaneous breaking of chiral 
symmetry,
the quark condensate $\Sigma $, in the thermodynamic and chiral limits
($V$ here is the size of the box). It is expected on the basis of lattice 
QCD simulations that chiral symmetry will be restored above some critical
temperature and/or chemical potential. 
By studying the Dirac spectrum in the infrared limit,
one hopes to obtain analytical information about the chiral phase transition. 

The presence of gauge fields alters the distribution of the Dirac
eigenvalues. For free quarks in 4d the spacing between eigenvalues is  
$\sim V^{-1/4}$, whereas for interacting quarks it is
$\sim V^{-1}$. Moreover, it is known that the Dirac eigenvalues
derived from QCD in a finite volume are constrained by sum rules 
first discovered by Leutwyler and Smilga \cite{LS}. These 
sum rules are obtained by
expanding the partition function in powers of the quark mass $m$
before and after averaging over the gauge field configurations,
and matching powers of $m$. 

The same sum rules 
can also be obtained from a random matrix theory with the same global 
symmetries as the QCD partition function. This was first noticed in 
\cite{V36}. In the random matrix theory (where the size of the 
random matrices $N$ is identified with the space-time volume $V$),
the average over gauge field configurations is substituted by an
average over Gaussian distributed random matrices with the same
structure of matrix elements as the Euclidean space 
Dirac operator $\gamma_\mu D_\mu$. The matrix elements can be real,
complex or quaternion real, corresponding to Dyson indices
$\beta =1$, $2$, $4$ respectively. The corresponding matrix models are
called the (chiral) Gaussian orthogonal, unitary and symplectic ensembles.

In fact, the extreme infrared limit of the QCD partition function
maps onto the same effective  
partition function (usually called the finite volume partition function)
as the random matrix theory 
\cite{V36,V26,V21,V31,V25,P4-P8,QCD3}. In retrospect this is not too
surprising, since the chiral Lagrangian to lowest order in the quark momenta
is completely determined by the pattern of chiral symmetry breaking 
and Lorentz invariance. In the so-called mesoscopic range 
$\Lambda_{QCD}^{-1} \ll L \ll \lambda_\pi^C$, where $L$ is the side of the box
and $\lambda_\pi^C$ is the Compton wavelength of the Goldstone modes,
this partition function expresses the quark mass
dependence in the static limit and in a finite volume. 
It is expressed as an integral over the 
Goldstone manifold of the composite variables corresponding to the pion fields
and it is a function
of one scaling variable $N\Sigma {\cal M}$, where ${\cal M}$ is a mass
matrix in flavor space. 
In substituting the complicated average over gauge field configurations with an
average over random matrices, we have the advantage that we can 
utilize a well-established mathematical framework that had its beginnings 
with the pioneering works in the field by Wigner, Dyson and Mehta. 

The sum rules can be expressed using the so-called microscopic
spectral density $\rho_S(\lambda)$, defined by
magnifying the distribution of eigenvalues in the vicinity of the
origin ($\lambda=0$) on the scale of the average eigenvalue spacing.
$\rho_S(\lambda)$ is a highly universal quantity 
\cite{V36,V31,V27-29,V33,V17V15,univers}. It is completely determined
by the global symmetries of the partition function, and does not depend
on the matrix potential, for example.

However, the sum rules are not sufficient for determining the spectral 
density itself. Fortunately, the Dirac spectrum can be obtained using the
so called partially quenched partition function for QCD \cite{OTV}, in
which one introduces bosonic and fermionic valence quark species 
in addition to the ordinary quarks. When the bosonic and fermionic 
valence quark masses coincide the partially quenched partition function
coincides with the original QCD partition function. For a nice review 
of this approach see also \cite{Vln}.

The cases corresponding to fundamental fermions with gauge group $SU(2)$, 
fundamental fermions with gauge group $SU(N_c)$
($N_c \ge 3$), and adjoint fermions with gauge group 
$SU(N_c)$ ($N_c \ge 2$) (labeled respectively by
$\beta =1$, $2$, $4$) in four dimensions, and the 
corresponding cases labeled by $\beta =1$, $2$ in three dimensions
have been analyzed in \cite{V21,V25,QCD3}. 
In this paper we will treat the only ``missing'' case, namely three-dimensional
QCD in the adjoint representation. 
We take the color gauge group to be
$SU(N_c)$ with $N_c\ge2$. This case will correspond to a matrix model 
labeled by $\beta =4$, as we will find that the Dirac operator is quaternion
real, that is, that its matrix elements can be written in the form
$Q_{kl}= a^0_{kl} 1 + i \vec a_{kl} \cdot \vec \sigma $ where the $a^\mu_{kl}$
are real numbers, $1$ is the $2\times 2$ unit matrix and $\vec \sigma $ 
the triplet of Pauli matrices. 
Like for all the other cases, we will propose a Gaussian
random matrix theory corresponding to the low-energy partition function
by substituting the integral over gauge fields with an integral over a
random, quaternion real matrix. By expressing
the fermion determinant as an integral over Grassmann
variables (utilizing the supersymmetric formalism developed in \cite{VWZ}), 
we will then perform the Gaussian integration corresponding in the field theory
to the average
over gauge fields. After manipulating the partition function further and 
eventually performing the Grassmann integration, we will obtain the finite
volume partition function. From this partition function the pattern of
spontaneous flavor symmetry breaking (assuming that such breaking takes place)
will emerge. We will assume from the outset 
that there is a nonzero condensate $\Sigma $.
The discrete parity symmetry discussed below and in \cite{V25,QCD3,Redlich} 
remains unbroken. (For a more detailed discussion of this issue see  
these references.)

\section{The parity-invariant Dirac operator in three dimensions}
\label{sec-Diracop}

To begin, let us consider
the Minkowski space Lagrangian for QCD in the adjoint representation

\begin{equation}
{\cal L} = -\frac{1}{4} {\rm tr} F^2 + \sum_{f=1}^{2N_f} \bar\psi_f 
(i\!\not{\!\! D} - m_f) \psi_f 
\end{equation}

where $F$ is the gauge field tensor, $\not{\! \! \! D} \equiv 
\gamma^\mu D_\mu$, $D_\mu$ is the covariant derivative for the adjoint
representation 
given explicitly in (\ref{eq:D_adjoint}), and $m_f$ is the quark mass 
corresponding to flavor $f$. For reasons that will become evident, we 
call the total number of flavors $2N_f$.
$\psi_f$ are quark spinors in the adjoint representation and $f$ is the 
flavor index (the indices corresponding to color and spin are suppressed). 

The lowest-dimensional representation of $\gamma^\mu$ is 
given by the Pauli matrices $\gamma^0=\sigma_3$, $\gamma^1=i\sigma_1$, 
$\gamma^2=i\sigma_2$. In this 2d representation, there is no chiral 
symmetry, since there is no $2\times 2$ matrix that anticommutes with the 
$\sigma_k$. 

For zero masses $m_f$, the above Lagrangian is invariant under parity $P$, 
defined in three dimensions by    

\bea
\label{eq:P}
\psi(t,x_1,x_2) &\to& \gamma_1 \psi(t,-x_1,x_2) \nonumber \\
A_0(t,x_1,x_2) &\to& A_0(t,-x_1,x_2)   \nonumber \\
A_1(t,x_1,x_2) &\to& -A_1(t,-x_1,x_2)   \nonumber \\
A_2(t,x_1,x_2) &\to& A_2(t,-x_1,x_2)    \nonumber \\
\eea

The mass term breaks this $P$ invariance. However, by choosing half of the 
masses equal to $+m$ and half equal to $-m$, we can achieve a  
$(P,Z_2)$--invariant Lagrangian \cite{QCD3,jeschris}.
In terms of 2--spinors, this choice corresponds to $N_f$ 
2--spinors $\phi_f$ with mass $+m$, and $N_f$ 2--spinors $\chi_f$ 
with mass $-m$.
Under $P$ the mass terms for the 2--spinors change sign, so that if the 
two sets of two-spinors transform into each other in a $Z_2$ transformation
$\phi_f \leftrightarrow \chi_f$, $f=1,2,...,N_f$, 
the total Lagrangian is invariant under the
combined transformations $P$ and $Z_2$. 
We can use this choice of mass term to write down a $(P,Z_2)$--invariant 
Lagrangian in the adjoint representation:

\beq
\label{eq:masschoice}
{\cal L} = -\frac{1}{4} {\rm tr} F^2 + \sum_{f=1}^{2N_f} \bar\psi_f 
i\!\not{\!\! D} \psi_f  - \sum_{f=1}^{N_f} m\bar\psi_f \psi_f
\, +\! \! \! \! \! \sum_{f=N_f+1}^{2N_f}\! \! \! \! m\bar\psi_f \psi_f
\eeq

As we will see in section \ref{sec-3dZ}, 
the given representation of $D_\mu $ 
-- in this case the adjoint representation -- uniquely
defines (in Euclidean space) the anti-unitary operator $Q$ that 
commutes with the Dirac operator $i\!\not{\! \! D}$. We will see that for the
adjoint representation, the condition $Q\psi =\psi $ 
leads in the fermionic partition function 
to an integral over only {\it half} of the
number of fermionic degrees of freedom with respect to the fundamental
representation. The condition $Q\psi =\psi $ is therefore called the Majorana
condition.

Also for $\beta =1$ (fundamental fermions with $SU(2)$ color
symmetry) we have an anti-unitary symmetry $[i\! \not{\! \! D},\tilde{Q}]=0$
(with the difference that $\tilde{Q}^2=+1$, whereas in the adjoint case we 
have $Q^2=-1$; see below), which leads to the same kind of
relation $\tilde{Q}\psi =\psi $. In the case of the fundamental 
$SU(2)$ representation
this relation defines a basis in which the Dirac operator has real matrix 
elements. In the adjoint case, because $Q^2=-1$, we will find that we can write
the Dirac operator in terms of real quaternions (see section \ref{sec-quat}).
In contrast to the adjoint case, in 
the fundamental $SU(2)$ case we integrate in the partition function
over both $\bar\psi $ and $\psi $, which are independent degrees of freedom.
This was shown explicitly in section 3 of ref.~\cite{QCD3}.
For the gauge groups $SU(N_c)$ ($N_c \ge 3$) and fundamental 
fermions, we have no anti-unitary symmetry.

To discuss the theory (\ref{eq:masschoice})
in Euclidean space, we first discuss the general issue 
of defining Majorana fermions in Euclidean space.

\section{Defining Majorana fermions in Euclidean space}
\label{sec-Majorana}

It is not in general
straightforward to define Majorana fermions 
in Euclidean space. In 4d, this is due to the fact that the would-be Majorana 
fermions do not transform like Dirac spinors under Euclidean 
Lorentz-transformations. This in turn is due to the absence, in Euclidean 
space, of an equation relating the right- and lefthanded 
Lorentz-transformations. For Minkowski space there is such a relation:

\beq
\sigma_2 \Lambda_R \sigma_2 = \Lambda_L^*
\eeq

This relation guarantees, in 4d Minkowski space, that the upper and
lower components of the usual Majorana fermions, $\xi_L $ and
$-\sigma_2 \xi_L^* $ transform like lefthanded and righthanded components,
respectively.  
In \cite{V21} it was shown that one can nevertheless 
construct a fermionic partition function for Majorana fermions in 4d 
(or 2d) Euclidean 
space. To begin with, we will here briefly recall how this was done.

Given the Euclidean Dirac operator for adjoint fermions, $i{\not{\! \! D}}$
(defined as usual by  ${\not{\! \! D}}=\gamma_\mu D_\mu $ where $\gamma_\mu $
are Euclidean gamma matrices satisfying 
$\{ \gamma_\mu ,\gamma_\nu \} = 2\delta_{\mu \nu }$ and 
$D_{\mu \, ab} = \partial_\mu \delta_{ab} + f_{abc} A_\mu^c $),
an anti-unitary operator $Q$ defined by

\beq
\label{eq:antiunitaryQ}
[i\! \not{\! \! D},Q]=0
\eeq

was identified. Since 
in 4d it turns out that $Q^2=-1$, the Majorana condition                       

\beq
\label{eq:Majcond}
Q\psi = \psi
\eeq

is contradictory (since it implies that $\psi = -\psi $ and thus $\psi =0$), 
unless we define

\beq
\psi^{**} = -\psi
\eeq

This condition is called conjugation of the second kind. It is 
common in the literature on Grassmann variables and in 
calculations involving supersymmetric random matrix theories.  
If $Q$ commutes with Euclidean Lorentz-transformations,
a partition function for Euclidean Majorana fermions 
can now be defined if we can find an operator ${\cal O}$ such that 

\beq
\label{eq:ZMaj}
Z=\sqrt{{\rm det}(i\! \not{\! \! D})}
= \int \! D\psi \, {\rm e}^{-\psi^T \! {\cal O} \, i\, {\not{D}} \psi}
\eeq

where $\psi $ are Majorana fermions satisfying $Q\psi =\psi $, 
${\rm det}\,{\cal O}=1$ and ${\cal O}i{\not{\! \! D}}$ is 
an antisymmetric operator. This last condition guarantees that the square 
root is well-defined. 
Lorentz-invariance of this partition function is guaranteed because 
in addition, we demand that ${\cal O}$ by construction satisfies 

\beq
\label{eq:Ocond}
\psi^T{\cal O} =\psi^\dagger \equiv (Q\psi )^\dagger
\eeq
 
(we recall that in Euclidean space the Lorentz-invariant quantity is
$\psi^\dagger \psi $). 
In (\ref{eq:ZMaj}) we integrate over only half of the
degrees of freedom with respect to the partition function for the 
usual Dirac fermions. 

Conversely, the Majorana condition can be identified by the condition
(\ref{eq:Ocond}), once we have written down the partition function,
by demanding Lorentz-invariance of the latter.

\section{The fermion determinant}
\label{sec-3dZ}

Following these hints, we can now construct a partition function for 
Euclidean Majorana fermions in 3d. As it turns out, the problems with 
defining Euclidean Majorana fermions that we encounter in four dimensions,
are not present in 3d because the Majorana fermions defined by
(\ref{eq:Majcond}) transform correctly
under 3d Euclidean Lorentz-transformations. This is because in 3d, there is
just one kind of 2-spinor and just one two-dimensional 
representation of the Lorentz group,
which is equivalent to the rotation group.
 
We start by defining Hermitian gamma matrices for our 3d Euclidean space. They
can simply be taken to be the Pauli matrices

\beq
\label{eq:gamma_mu}
\gamma_0=\sigma_3, \ \ \ \ \ \gamma_1=\sigma_1, \ \ \ \ \ \gamma_2=\sigma_2
\eeq

satisfying 
$\{ \gamma_\mu ,\gamma_\nu \} = 2\delta_{\mu \nu }$. The Dirac
operator in the adjoint representation is 

\beq
\label{eq:D_adjoint}
\gamma_\mu D_{\mu\, ab} = \gamma_\mu (\partial_\mu \delta_{ab} +
i A_\mu^c (T_c)_{ab}) = 
\gamma_\mu (\partial_\mu \delta_{ab} + f_{abc} A_\mu^c ) 
\eeq

where $f_{abc}$ are real structure constants for the gauge group. 
$D_\mu $ is antisymmetric 
under transposition and ${\not{\! \! D}}$ is antihermitean. It is easy to
find the antiunitary operator that satisfies (\ref{eq:antiunitaryQ})
in this case. It
is given by $Q=i\gamma_2 K$ 
where $i\gamma_2 \equiv C$ is the charge conjugation 
operator satisfying $C\gamma_\mu^*C^{-1} = -\gamma_\mu $ and $K$ denotes 
complex conjugation. We see that like in the 4d case outlined above,
$Q^2=-1$ and the Majorana condition 

\beq
\label{eq:MajcondCK}
\psi = CK\psi
\eeq

makes sense only 
if we introduce conjugation of the second kind, $\psi^{**}=-\psi $.
The Majorana condition is then consistent with the explicit form

\beq
\psi = \left( \begin{array}{c} \chi \\ -\chi^* \end{array} \right)
\eeq

of the spinors. 
The generators of the Euclidean Lorentz-group corresponding to the 
representation
(\ref{eq:gamma_mu}) of $\gamma_\mu $ are given by \cite{Peskin}

\beq
S^{\mu \nu }=\frac{i}{4} [\gamma_\mu ,\gamma_\nu ]
\eeq

We find $S_{01}=-\sigma_2/2$, $S_{02}=\sigma_1/2$ and $S_{12}=-\sigma_3/2$.
These are equivalent to the generators of the rotation group and one can
easily verify that the corresponding 
Lorentz-transformations commute with the anti-unitary 
operator $Q$ defining the Majorana condition in our case. 

To write down the fermionic action we now look for an operator ${\cal O}$
such that ${\rm det}\,{\cal O}=1$ and $\psi^T {\cal O} =\psi^\dagger =
(Q\psi )^\dagger $. We immediately arrive at

\beq
{\cal O} = -i\sigma_2
\eeq

In addition, ${\cal O}i{\not{\! \! D}}$ is antisymmetric so
it is now straightforward to write down a partition function for 
Majorana fermions satisfying (\ref{eq:MajcondCK}) in 3d Euclidean space

\beq
\sqrt{{\rm det}(i\not{\! \! \! D})} = 
\sqrt{{\rm det}({\cal O}i\not{\! \! \! D})}=
\int \! D\psi \, {\rm e}^{-\psi^T \! {\cal O}\, i\, {\not{D}} \psi}
\eeq

\section{The Dirac operator for Majorana fermions}
\label{sec-quat}

In analogy with 4d, we expect that there is a basis in which 
$i{\not{\! \! D}}$ 
can be written in the form of real quaternions \cite{V21}. 
We will now verify this. It is easily checked (using 
$C^2 = -1$, $\chi_k^{**} = -\chi_k $ and $\hat \phi_k^{**} = +\hat \phi_k $) 
that an expansion of $\psi $ consistent with the Majorana condition is 

\beq
\psi = \sum_k\left( \hat \phi_k \chi_k + C\hat \phi_k^* \chi_k^* \right)
\eeq

where $\hat \phi_k$ are arbitrary c-number 2-spinors and $\chi_k$ are 
Grassmann variables. 
Since $(CK)^2=-1$, $\hat \phi_k$ and $C\hat \phi_k^*$ are linearly independent.

We then see that the expression in the fermionic action involving the 
Dirac operator can be written

\bea
\label{eq:Qform}
\psi^\dagger i\! \not{\! \! D} \psi 
& = & \sum_{kl} \left( \hat \phi_k^\dagger \chi_k^* - 
\left( C\hat \phi_k^* \right)^\dagger \! \! \chi_k \right) i\! \not{\! \! D}
\left( \hat \phi_l \chi_l + C\hat \phi_l^* \chi_l^* \right) \cr
& = &\sum_{kl} \left( \begin{array}{c} \chi_k \\ \chi_k^* \end{array} \right)^*
\left( \begin{array}{cc} \hat\phi_k^\dagger i\! \not{\! \! D} \hat\phi_l &     
                         \hat\phi_k^\dagger i\! \not{\! \! D} C\hat\phi_l^* \\
                         \hat\phi_k^TC^\dagger i\! \not{\! \! D} \hat\phi_l &
                         \hat\phi_k^TC^\dagger i\! \not{\! \! D} C\hat\phi_l^*
       \end{array} \right)
 \left( \begin{array}{c} \chi_l \\ \chi_l^* \end{array} \right)
\eea

The matrix 

\beq
\label{eq:Q_kl}
Q_{kl} \equiv 
\left( \begin{array}{cc} \hat\phi_k^\dagger i\! \not{\! \! D} \hat\phi_l &     
                         \hat\phi_k^\dagger i\! \not{\! \! D} C\hat\phi_l^* \\
                         \hat\phi_k^TC^\dagger i\! \not{\! \! D} \hat\phi_l &
                         \hat\phi_k^TC^\dagger i\! \not{\! \! D} C\hat\phi_l^*
       \end{array} \right)
\eeq

has the form 

\beq
Q_{kl}=\left( \begin{array}{cc} A & B \\ -B^* & A^* \end{array} \right)
\eeq

This can be verified using the properties of $C$ and 
eq.~(\ref{eq:antiunitaryQ}) from which we derive

\bea
C^\dagger i\! \not{\! \! D} C = (i\! \not{\! \! D})^* \cr
C^\dagger i\! \not{\! \! D} = - (i\! \not{\! \! D})^* C
\eea

$Q_{kl}$ can be written as a real quaternion: 

\beq
Q_{kl}= a^0_{kl} 1 + i \vec a_{kl} \cdot \vec \sigma 
\eeq

where $a^0_{kl}$, $a^i_{kl}$ are real numbers. 

\section{Random matrix theory}
\label{sec-RMT}

We are now ready to define the random matrix theory for massive, 
parity-invariant three-dimensional QCD 
with fermions in the adjoint representation. 
In analogy with all the other cases in 3d 
\cite{V25,QCD3} and in 4d \cite{V36,V26,V21,V27-29}, we substitute the integral
over gauge field configurations in the partition function for QCD with an
integral over a Hermitian random matrix $T$. We take this matrix to have the 
quaternion structure (\ref{eq:Q_kl}). Using the same notation as in \cite{V21}
we therefore set 

\beq
\label{eq:quaternion_structure}
T_{ij}=\sum_{\mu =0}^3 a^{\mu }_{ij} i\bar{\sigma }_\mu
\eeq

where $a^{\mu }_{ij}$ are real numbers and
we have defined $\bar{\sigma }_0=-i$ and $\bar{\sigma}_1$, $\bar{\sigma}_2$, 
$\bar{\sigma}_3$ are the usual Pauli matrices. 
This means the matrix elements of 
$T$ are themselves $2\times 2$ matrices. The operator ${\not{\! \! D}}$ 
is antihermitian, so we can 
substitute ${\not{\! \! D}}$ in the Euclidean fermion determinant 
with the matrix $iT$ where $T$ is Hermitian:

\bea
\label{eq:RMT}
Z(m_1,...,m_{2N_f}) &=& \int DA\, {\rm e}^{-S[A]} \prod_{f=1}^{2N_f}
\sqrt{{\rm det}(\not{\!\! D} + m_f)} \cr
&\to &
\int DT\, {\rm e}^{-N\Sigma^2 {\rm tr}(T^\dagger T)} \prod_{f=1}^{2N_f}
\sqrt{{\rm det}(iT+m_f)}
\eea

This random matrix theory has the same global flavor symmetry as the
QCD partition function. In addition,
$T$ is here taken to be a matrix of $N \times N$ real quaternions, so that
the anti-unitary symmetry of adjoint QCD is reproduced, and as we will see,
the flavor symmetry breaking pattern will appear from the random matrix
partition function. 
$DT$ is the invariant (Haar) measure. As usual 
\cite{V36,V26,V21,V25,QCD3,V27-29} $N$ is identified with the space-time 
volume. We will see that $\Sigma$ is the value of the
condensate (the order parameter for spontaneous symmetry breaking)
defined by 

\beq
\Sigma = -\lim_{m_f \to 0} \lim_{N \to \infty} \frac{1}{N}\frac{\partial}
{\partial m_f} \ln Z(m_1,...,m_{2N_f})
\eeq

We are assuming that this condensate is non-zero.
As discussed in Section \ref{sec-Diracop}, we choose the masses in 
pairs of opposite sign, so that the corresponding Minkowski space
theory is parity-invariant. We call the total number of flavors $2N_f$,
and we choose $m_f=+m$ for $f=1,...,N_f$ and $m_f=-m$ for $f=N_f+1,...,2N_f$,
where $m$ is a real positive number. 

In order to evaluate the partition function (\ref{eq:RMT}) we rewrite the
square root of the fermion determinant as an integral over Grassmann
variables (cf. (\ref{eq:Qform})):

\beq
\label{eq:det}
\prod_f \sqrt{\det(iT+m_f)}
= \int \prod_f D\phi_f {\rm exp}\left[ -i\sum_f {\phi^i_f}^* 
(T-im_f)_{ij} \phi^j_f \right]
\eeq

where the indices $i$, $j$ are summed over 
from $1$ to $N$. Here we are 
using the formalism developed in 
\cite{VWZ} for supersymmetric matrix integrals.
Our matrix integral is pure fermionic and involves no bosonic variables,
so we will apply this formalism using only the fermion-fermion block. Our
integration measure is  

\begin{equation}
\prod_f D\phi_f = \prod_{f=1}^{2N_f} \prod_{i=1}^N 
d{\chi^i_f}^* d\chi^i_f 
\end{equation}

Recall that $T_{ij}$ are quaternions, so the $\phi^i_f$ are 2-component
vectors like in (\ref{eq:Qform}):

\beq
\phi^i_f=\left( \begin{array}{l} \chi^i_f \\ {\chi^i_f}^* \end{array} \right)
\eeq
 
We use conjugation of the second kind $\chi^{**}=-\chi$ for the Grassmann 
variables.

The first step is to perform the Gaussian integral over the random matrix $T$
in the integral 

\bea
\label{eq:1step}
Z(m)&=&
\int DT \int \prod_f D\phi_f 
{\rm exp}\left[ -N\Sigma^2 {\rm tr}(T^\dagger T) -i\sum_f {\phi_f}^* 
(T-im_f) \phi_f \right] \cr
&=& \int Da^\mu \prod_f D\phi_f {\rm exp}\left[ -N\Sigma^2 a^{\mu }_{ij}
a^{\mu }_{ij}+ \sum_f {\phi^i_f}^*a^{\mu }_{ij}\bar{\sigma}_\mu \phi^j_f
-\sum_f m_f {\phi^i_f}^*\phi^i_f \right]
\eea

where we sum over repeated indices $\mu $ and $i$, $j$. In the second step
of (\ref{eq:1step}) we have used that $\bar\sigma_\mu {\bar\sigma_\nu}^\dagger
+\bar\sigma_\nu {\bar\sigma_\mu}^\dagger =2\delta_{\mu \nu}$.
To perform the Gaussian integral we complete the
square in the exponent of (\ref{eq:1step}) by setting

\beq
\label{eq:complete_square}
a^{\mu }_{ij} \to  \tilde{a}^{\mu }_{ij} 
\equiv  a^{\mu }_{ij}
- \frac{1}{2\Sigma^2 N} \sum_f {\phi^i_f}^* \bar{\sigma}_\mu \phi^j_f
\eeq

The symmetry properties of the $\tilde{a}^{\mu }_{ij}$ are the same as 
the symmetry properties of $a^{\mu }_{ij}$, namely $a^0_{ij}=a^0_{ji}$,
$a^k_{ij}=-a^k_{ji}$ ($k=1,2,3$). These symmetry properties follow from
the hermiticity of the matrix $T$ and (\ref{eq:quaternion_structure}). 
The integral over $T$ is uniformly convergent in the fermionic variables
$\phi^i_f$, so we can interchange the two integrals and perform the Gaussian
integration. We then arrive at

\beq
Z(m) \sim  \int \prod_f D\phi_f {\rm exp}\left[ \frac{1}{4\Sigma^2 N}
\sum_{fg} {\phi^i_{f}}^* \bar\sigma_\mu \phi^j_f \,
{\phi^i_g}^* \bar\sigma_\mu \phi^j_g -\sum_f m_f {\phi^i_f}^*\phi^i_f \right]
\eeq

To evaluate $\sum_{\mu} {\phi^i_{f}}^* \bar\sigma_\mu \phi^j_f \, 
{\phi^i_g}^* \bar\sigma_\mu \phi^j_g$ we use the Fierz' identity
\cite{V21} 

\beq
\sum_\mu \bar\sigma_\mu^{\alpha\beta}\bar\sigma_\mu^{\gamma\delta}
=2(\delta_{\alpha\delta}\delta_{\gamma\beta}-\delta_{\alpha\beta}
\delta_{\gamma\delta})
\eeq

and insert the explicit form for the 2-component Grassmann variables
$\phi^i_f$

\beq
\phi^i_f=\left( \begin{array}{l} \chi^i_f \\ {\chi^i_f}^* \end{array} \right)
\eeq

to arrive at

\beq
\sum_{\mu} {\phi^i_{f}}^* \bar\sigma_\mu \phi^j_f 
{\phi^i_g}^* \bar\sigma_\mu \phi^j_g = 2F^2_{fg}
\eeq

where we have set 

\beq
F_{fg}\equiv {\chi^i_f}^*\chi^i_g + {\chi^i_g}^*\chi^i_f
\eeq

(a sum over $i$ is understood). Disregarding for a moment the mass term,
we have rewritten the exponent in $Z(m)$ as a square. We can therefore
use the Hubbard-Stratonovitch transformation \cite{VWZ}

\beq
\exp \left[-\alpha F_{fg}F_{fg}\right] \sim \int d\sigma\!_{fg} 
\exp \left[-\frac{1}{4\alpha} \sigma_{fg}\sigma_{fg} -i \sigma_{fg} F_{fg}
\right]
\eeq

where $\sigma_{fg}$ is a real variable, to rewrite the partition function  
as

\beq
\label{eq:afterHubbard}
Z(m) \sim  \int \prod_f D\chi_f D\sigma \, {\rm exp}\left[ -\frac{\Sigma^2 N}{2}
\sigma^2_{fg} +\sigma_{fg}F_{fg} - 2\sum_f m_f {\chi^i_f}^*\chi^i_f \right] 
\eeq

In (\ref{eq:afterHubbard}), $D\sigma$ is the Haar measure for 
the real symmetric matrix $\sigma $. In order to preserve the flavor 
symmetry of $Z(m)$, $\sigma $ is chosen symmetric like the matrix $F$.
Interchanging again the Grassmann integral with the integral over 
$\sigma $, and 
writing out the Grassmann components of $F_{fg}$ we see that
the partition function is proportional to

\beq
Z({\cal M}) 
\sim \int D\sigma \prod_f D\chi_f \, {\rm exp}\left[ -\frac{\Sigma^2 N}{2}
{\rm tr}\left( \sigma \sigma^T \right) + 2{\chi^i}^*(\sigma+{\cal M})\chi^i
\right]
\eeq

where ${\cal M}$ is the $2N_f \times 2N_f $ mass matrix in flavor space,

\beq
\label{eq:calM}
{\cal M} = \left( \begin{array}{cccccc} 
            m &        &    &    &        &      \\
              & \ddots &    &    &        &      \\
              &        &  m &    &        &      \\
              &        &    & -m &        &      \\
              &        &    &    & \ddots &      \\ 
              &        &    &    &        &  -m  \\
\end{array} \right)
\eeq

Performing the Grassmann integrals we get 

\beq
\label{eq:Z(S)}
Z({\cal M}) \sim \int DS \, {\rm e}^{-\frac{\Sigma^2 N}{2} {\rm tr}(SS^T)} 
{\rm det}^N(S+{\cal M})
\eeq

with $S$ a symmetric real matrix and $DS$ the Haar measure. 
In the next section we will evaluate
this expression using a saddle point analysis to find the low-energy
effective partition function for three-dimensional QCD with adjoint fermions. 

\section{The effective partition function}
\label{sec-Z_eff} 

We now decompose the real symmetric matrix $S$ in (\ref{eq:Z(S)}) into 
``polar'' coordinates \cite{Hua}:

\beq
\label{eq:polar}
S=O\Lambda O^T
\eeq

where $O$ is a real orthogonal matrix and $\Lambda $ is
the real diagonal matrix
 
\beq
\label{eq:Lambda}
\Lambda = \left( \begin{array}{ccc}
                 \lambda_1 &        &                \\
                           & \ddots &                \\
                           &        & \lambda_{2N_f} \end{array} \right),
\eeq

One can always choose the polar coordinates such that 

\beq
\label{eq:lambda_cond}
\lambda_1 \ge \lambda_2 \ge ... \ge \lambda_{2N_f}
\eeq

The integration variables in $S$ and those in $O\Lambda O^T$ will be in 
one-to-one correspondence if the integral over $DS$ is taken to be

\beq
\label{eq:polarmeas}
\int DS = \int_{O \in [O]} DO D\Lambda J(\Lambda )
\eeq

where $[O]$ denotes the set of left cosets of the group of 
$2N_f \times 2N_f $ real orthogonal 
matrices with respect to the subgroup consisting of matrices of the form

\beq
\left( \begin{array}{cccc}
          \pm 1 &       &     &   \\
                & \pm 1 &     &   \\
                &       & \ddots &    \\
                &       &        & \pm 1 \end{array} \right)
\eeq
  
and $J(\Lambda )$ is the Jacobian corresponding to (\ref{eq:polar}).
This Jacobian was given in \cite{Hua} and is proportional to

\beq
J(\Lambda ) \propto \prod_{i<j} |\lambda_i - \lambda_j |
\eeq

We are now ready to 
determine the saddle point of the partition function at 
zero mass. Setting ${\cal M}=0$ in (\ref{eq:Z(S)}) and making the substitution 
(\ref{eq:polar}), (\ref{eq:polarmeas}) in $Z(0)$ the partition function takes
the form

\beq
Z(0) \sim \int D\Lambda {\rm exp}\left[ -\frac{\Sigma^2 N}{2} \sum_f
\lambda_f^2 + N \sum_f {\rm ln} \lambda_f 
- {\rm ln} J(\lambda_1,...\lambda_{2N_f} )  \right]
\eeq

The matrix $\Lambda $ is diagonal and the saddle point is given by

\beq
\lambda_f = \pm \frac{1}{|\Sigma |}
\eeq

since the Jacobian drops out in the large $N$ limit. 
It follows from the derivation of the Banks-Casher formula that if the    
flavor symmetry is broken spontaneously, the condensate
for each flavor has to have the same sign as the mass \cite{BC,V25,QCD3},
so we choose, in accordance with (\ref{eq:lambda_cond}) and (\ref{eq:calM})

\beq
\label{eq:Lambda_sp}
\Lambda_{sp} = \frac{1}{|\Sigma |} \left( \begin{array}{cccccc} 
            1 &        &    &    &        &      \\
              & \ddots &    &    &        &      \\
              &        &  1 &    &        &      \\
              &        &    & -1 &        &      \\
              &        &    &    & \ddots &      \\ 
              &        &    &    &        &  -1  \\
\end{array} \right) \equiv \frac{1}{|\Sigma |} J
\eeq

We now expand the determinant in (\ref{eq:Z(S)}) at the saddle point,
$\Lambda = \Lambda_{sp} $ for a small mass matrix ${\cal M}\neq 0$. To
first order in ${\cal M}$ we find

\bea
Z({\cal M}) &\sim & \int DO \, {\rm det}^N(O\Lambda_{sp} O^T + {\cal M}) \cr
     &\sim & \int DO \, {\rm det}^N(O\Lambda_{sp} O^T) {\rm e}^{N{\rm tr}\, 
{\rm ln}(1+O\Lambda_{sp}^{-1} O^T {\cal M}})\cr
     &\propto & \int DO \, {\rm e}^{N\Sigma {\rm tr}(OJO^T {\cal M})}
\eea

The matrix $J$ was defined in (\ref{eq:Lambda_sp}). This is our expression for 
the low-energy effective partition function. As usual, it is a function of the
scaling variable $N\Sigma {\cal M}$. The matrix $J$ is invariant under the 
subgroup $O(N_f)\times O(N_f)$, so in our final expression for $Z({\cal M})$,

\beq
\label{eq:final}
Z({\cal M}) \sim  \int_{O(2N_f)/(O(N_f)\times O(N_f))} 
DO \, {\rm e}^{N\Sigma {\rm tr}(OJO^T {\cal M})}
\eeq

the integral goes over the coset space $O(2N_f)/(O(N_f)\times O(N_f))$. 
The flavor symmetry breaking pattern is thus $O(2N_f) \to O(N_f)\times O(N_f)$.
This is also what we expect from the fermionic action,

\beq
S_F = \int d^3x \sum_{f=1}^{2N_f} 
\psi_f^T {\cal O}(\not{\!\! D} + m_f) \psi_f
\eeq

It is evident that 
$\sum_f \psi_f^T {\cal O}\! \not{\!\! D}\psi $ is invariant under $O(2N_f)$
transformations in flavor space (note that the operator 
${\cal O}\! \not{\!\! D}$
is diagonal in flavor space), while the mass term $\sum_f m_f \psi_f^T \psi_f $
is invariant under $O(N_f)\times O(N_f)$, which is the unbroken subgroup.
The dimension of the coset is

\beq
\label{eq:dimcoset}
M=\frac{2N_f(2N_f-1)}{2}-2\frac{N_f(N_f-1)}{2}=N_f^2
\eeq

\section{Sum rules}

We can easily derive the first Leutwyler-Smilga like sum rule for the
eigenvalues of the Dirac operator. We will use the same method as in 
\cite{V21,QCD3}.  

The sum rules are obtained by expanding the expression for $Z({\cal M})$,
eq.~(\ref{eq:final}) and comparing the coefficients order by order in $m^2$
to the (normalized) expectation value of the fermion determinant:

\beq
\left\langle \prod_f \prod_{\lambda_k > 0} 
\left( 1+ \frac{m^2}{\lambda^2_k}\right) 
\right\rangle
\eeq

(Note that because of our choice of $(P,Z_2)$-invariant Lagrangian, 
there is an effective chiral symmetry that makes the spectrum symmetric
around $\lambda =0$.
This chiral symmetry applies to the $4\times 4$ gamma matrices given in
eq.~(2.4) of ref.~\cite{QCD3}:

\beq
\label{eq:gamma_mu}
\gamma^0 = \left( \begin{array}{cc} \sigma_3 & 0 \\ 0 & -\sigma_3 \end{array}
\right),\ \ \ \ \ 
\gamma^1 = \left( \begin{array}{cc} i\sigma_1 & 0 \\ 0 & -i\sigma_1 \end{array}
\right),\ \ \ \ \ 
\gamma^2 = \left( \begin{array}{cc} i\sigma_2 & 0 \\ 0 & -i\sigma_2 \end{array}
\right)
\eeq

or equivalently to the corresponding Euclidean gamma matrices. 
We could just as well write ${\cal L}$ in terms of these.) 
Here the expectation value is defined as

\beq
\langle f(\lambda,m) \rangle = \frac{\int DA \, {\rm e}^{-S[A]}\,(\prod_{k,f} 
{\lambda^k_f}^2) \, f(\lambda,m)}{\int DA \,  {\rm e}^{-S[A]}\, (\prod_{k,f} 
{\lambda^k_f}^2) \, f(\lambda,0)}
\eeq

where $A$ is the gauge field and $S[A]$ the Euclidean Yang-Mills action. 
Expanding the integrand in (\ref{eq:final}) the surviving group integrals 
at order $m^2$ have the form 

\beq
\label{eq:zeta(X)} 
\zeta (X) =\int_{O\in G/H} DO\, {\rm tr}(OJO^TX) {\rm tr}^*(OJO^TX) \equiv
\int_{O\in G/H} DO\, {\rm tr}^2(OJO^TX)
\eeq

where $G/H$ is the coset and $X \equiv N\Sigma {\cal M}\equiv N\Sigma mJ$. 
The first order term is killed by the group integration. We note that the
matrices $OJO^T$ are $2N_f \times 2N_f$ symmetric unimodular matrices.
We now choose real, symmetric and traceless generators $t_k$, $k=1,...,M_s$ 
for these. $M_s$ is the number of such generators:

\beq
M_s=\frac{2N_f(2N_f+1)}{2}-1
\eeq

The generators can be normalized as follows:

\beq
\label{eq:normalization}
{\rm tr}(t_kt_l)=2N_f \delta_{kl}
\eeq

This normalization is natural since the generators are $2N_f \times 2N_f$
matrices. We wish to
choose $t_1$ such that $t_1=J$. In doing this we can use
the relation 

\beq
\label{eq:traceformula}
\sum_{k=1}^{M_s} {\rm tr}(At_k){\rm tr}(Bt_k) = 2N_f {\rm tr}(AB)
\eeq

to calculate the value of $\zeta (X)$. Equation (\ref{eq:traceformula})
is valid for any two symmetric unimodular matrices $A$ and $B$. 
Because of the invariance of the measure
in the integral defining $\zeta (X)$, one finds that $\zeta (X)$ is 
proportional to ${\rm tr}(X^\dagger X)$. Due to eq~(\ref{eq:normalization})
and the choice of $t_1$, we find $\zeta (t_1)=\zeta (t_2)=
...=\zeta (t_M)$. Therefore we can set 

\beq
\zeta (t_1) = \frac{1}{M_s} \sum_{k=1}^{M_s} \int DO\, {\rm tr}^2(OJO^T t_k)
\eeq

Using (\ref{eq:traceformula}) and ${\rm tr}(J^2)=2N_f$ we now immediately 
see that

\beq
\zeta (X) =\frac{1}{M_s} vol(G/H) (N\Sigma m)^2 (2N_f)^2
\eeq

Inserting this into the expansion we get 

\beq
\frac{Z(m)}{Z(0)} =\left\langle 1+m^2 \, 2N_f \sum_{\lambda_k > 0} 
\frac{1}{\lambda^2_k} 
+ ... \right\rangle = 1+\frac{1}{2} (N\Sigma m)^2 \frac{1}{M_s} (2N_f)^2 + ... 
\eeq

where the volume of the coset cancels in the ratio. Inserting the value of 
$M_s$ we therefore arrive at the sum rule  

\beq
\left\langle \sum_{\lambda_k > 0} 
\frac{1}{(N\Sigma \lambda_k)^2} \right\rangle =\frac{2N_f}{2(2N_f-1)(N_f+1)}
\eeq

where $2N_f$ is the original number of flavors. Considering also the 
sum rules found in \cite{V25,QCD3} for Dyson index $\beta=2,\ 1$ in 3d,
the sum rules for three dimensions can be summarized as follows:

\begin{equation}
\left\langle \sum_{\lambda_k > 0} 
\frac{1}{(N\Sigma \lambda_k)^2} \right\rangle = \frac{2N_f}{2(2N_f-1)
(\frac{4N_f}{\beta }+1)}
\end{equation}

where $2N_f$ is the number of flavors and $\beta $ the Dyson index.
The corresponding formula for the 4d cases was given in \cite{V26,V21}.
In the published 
version of this paper, as well as in the previous electronic version,
the wrong sum rule was given due to an error in (\ref{eq:traceformula}).

\section{Conclusion}

We have obtained the mass dependence of the 
finite volume partition function for three-dimensional QCD 
with quarks in the adjoint representation and gauge group $SU(N_c)$ 
($N_c \ge 2$). We chose the quark masses such that the corresponding 
Minkowski space QCD Lagrangian is $(P,Z_2)$--invariant, where $P$ denotes 
the parity transformation defined in three dimensions by eq.~(\ref{eq:P}).

As a starting point we used a random 
matrix theory with the same global symmetries as the gauge theory. 
This is possible since the gauge theory and the random matrix theory 
are equivalent in the mesoscopic regime. 

The effective partition function describes the static limit
of the Goldstone modes resulting from the spontaneous breaking of
global flavor symmetry. It is determined by the symmetry
breaking pattern. We assumed that flavor
symmetry breaking occurs, and found that the global flavor symmetry $O(2N_f)$
is broken by the vacuum state to $O(N_f) \times O(N_f)$ symmetry.

Although the author does not claim to understand the 
classification of symmetric superspaces, it is interesting to 
note that the result obtained here for the Goldstone manifold, together 
with the other Goldstone manifolds obtained 
for $\beta =1$, $2$, and $4$ in 3d and 4d, exhaust the scheme of Zirnbauer 
\cite{Zirn} with respect to 
the fermionic symmetric spaces $M_F$. (I thank Poul Damgaard and Jac 
Verbaarschot for this remark.) We can see this by looking at Table~3 of 
\cite{Zirn} and matching the various random matrix ensembles with the 
fermionic symmetric spaces given by Zirnbauer. These spaces,  
together with the corresponding bosonic symmetric spaces $M_B$, make
up the Riemannian symmetric superspaces corresponding to 
the various supersymmetric generating
functions. In simple terms, each 
symmetric superspace corresponds to a random matrix theory.
For us the $M_F$ define the respective Goldstone manifolds, since we do not
have any bosonic sector. From the fermionic symmetric spaces given by
Zirnbauer \cite{Zirn} we confirm that our case,
the Gaussian symplectic ensemble, corresponds to the random matrix theory 
denoted AII in 
Table~3 of \cite{Zirn}, while the coset spaces obtained in  
\cite{V21,V25,QCD3} nicely fit the compact symmetric spaces 
belonging to the classes labelled by A, AI, AIII, BDI, and CII. 
The classes A, AI, and AII correspond to the non-chiral ensembles 
while AIII, BDI, and CII correspond to the chiral ones. 

\section{Acknowledgments}

I thank Poul Damgaard, Jac Verbaarschot, Christian Hilmoine and Rune Niclasen  
for discussions and Poul Damgaard for reading the manuscript. This research 
was supported by the European Union through a research grant of the 
category ``Marie Curie''.


\begin{thebibliography}{99}

\bibitem{BC} T. Banks and A. Casher, Nucl. Phys. B169 (1980) 103

\bibitem{LS} H. Leutwyler and A. Smilga, Phys. Rev. D46 (1992) 5607

\bibitem{V36} E.V. Shuryak and J.J.M. Verbaarschot, Nucl. Phys. A560 (1993) 306

\bibitem{V26} A. Smilga and J.J.M. Verbaarschot, Phys. Rev. D51 (1995) 829

\bibitem{V21} M.A. Halasz and J.J.M. Verbaarschot, Phys. Rev. D52 (1995) 2563

\bibitem{V31} J.J.M. Verbaarschot, Acta Phys. Polon. B25 (1994) 133

\bibitem{V25} J.J.M. Verbaarschot and I. Zahed, Phys. Rev. Lett. 73 (1994) 2288

\bibitem{P4-P8} P.H. Damgaard, Phys. Lett. B424 (1998) 32, G. Akemann and 
P.H. Damgaard, Nucl. Phys. B528 (1998) 411, 
Phys. Lett. B432 (1998) 390,  P.H. Damgaard, hep-th/9807026

\bibitem{QCD3} U. Magnea, Phys. Rev. D 61, 56005 (2000), hep-th/9907096 

\bibitem{V27-29} J.J.M. Verbaarschot, Phys. Rev. Lett. 72 (1994) 2531,
Nucl. Phys. B426 (1994) 559, J.J.M. Verbaarschot, hep-th/9405006

\bibitem{V33} J.J.M. Verbaarschot and I. Zahed, Phys. Rev. Lett. 70 (1993) 3852

\bibitem{V17V15} A.D. Jackson, M.K. \c Sener, J.J.M. Verbaarschot,
Nucl. Phys. B479 (1996) 707

\bibitem{univers} G. Akemann, P. H. Damgaard, U. Magnea and S. Nishigaki
Nucl. Phys. B487 (1997) 721

\bibitem{OTV} J. Osborn, D. Toublan and J. Verbaarschot, Nucl. Phys. B540 
(1999) 317

\bibitem{Vln} J.J.M. Verbaarschot, {\it ``The infrared limit of the QCD Dirac 
spectrum and applications of chiral random matrix theory to QCD''}
(lectures given at the APCTP-RCNP Joint International School on Physics of 
Hadrons and QCD, Osaka, 1998, and the 1998 YITP-Workshop on QCD and Hadron 
Physics, Kyoto, 1998), hep-ph/9902394

\bibitem{VWZ} J.J.M. Verbaarschot, H.A. Weidenm\"uller and M.R. Zirnbauer,
Phys. Rep. 129 (1985) 367

\bibitem{Redlich} A. N. Redlich, Phys. Rev. Lett. 52 (1984) 18, Phys. Rev. D29
(1984) 2366

\bibitem{jeschris} J. Christiansen, Nucl. Phys. B547 (1999) 329 and 
Master's thesis, {\it ``Universality of Random Matrix Theories of QCD''} 
(the Niels Bohr Institute, February 1999)

\bibitem{Peskin} M. E. Peskin and D. V. Schroeder, {\it ``An introduction to 
quantum field theory''}, Addison-Wesley 1995

\bibitem{Hua} L. Hua, {\it ``Harmonic Analysis of functions of several complex 
variables in the classical domain''}, American Mathematical Society (1963) 

\bibitem{Zirn} M. Zirnbauer, J. Math. Phys. 37 (1996) 4986

\end{thebibliography}
\end{document}